\documentclass[8.5pt,twoside,twocolumn]{article}
\usepackage[utf8]{inputenc}
\usepackage[T1]{fontenc}
\usepackage{graphicx}
\usepackage{tikz}
\usetikzlibrary{calc}
\usepackage{subfigure}
\usepackage{lmodern}
\usepackage{wrapfig}
\usepackage{float}
\usepackage{amsmath}
\usepackage{upgreek}
\usepackage{array}
\usepackage{color}
\usepackage{transparent}
\usepackage{multirow}
\usepackage{epstopdf}
\usepackage{amsmath,amsfonts,amssymb}

\usepackage[super,sort&compress,comma]{natbib} 
\usepackage{mhchem}
\usepackage{times,mathptmx}
\usepackage{sectsty}
\usepackage{balance} 

\usepackage{graphicx} 
\usepackage{lastpage}
\usepackage[format=plain,justification=raggedright,singlelinecheck=false,font=small,labelfont=bf,labelsep=space]{caption} 
\usepackage{fancyhdr}
\begin{document}

\thispagestyle{plain}
\fancypagestyle{plain}{
\fancyhead[L]{\includegraphics[height=8pt]{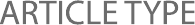}}
\fancyhead[C]{\hspace{-1cm}\includegraphics[height=20pt]{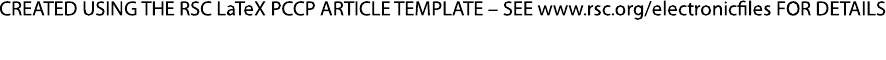}}
\fancyhead[R]{\includegraphics[height=10pt]{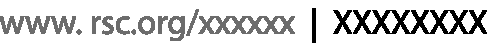}\vspace{-0.2cm}}
\renewcommand{\headrulewidth}{1pt}}
\renewcommand{\thefootnote}{\fnsymbol{footnote}}
\renewcommand\footnoterule{\vspace*{1pt}%
\hrule width 3.4in height 0.4pt \vspace*{5pt}} 
\setcounter{secnumdepth}{5}

\makeatletter 
\def\subsubsection{\@startsection{subsubsection}{3}{10pt}{-1.25ex plus -1ex minus -.1ex}{0ex plus 0ex}{\normalsize\bf}} 
\def\paragraph{\@startsection{paragraph}{4}{10pt}{-1.25ex plus -1ex minus -.1ex}{0ex plus 0ex}{\normalsize\textit}} 
\renewcommand\@biblabel[1]{#1}            
\renewcommand\@makefntext[1]%
{\noindent\makebox[0pt][r]{\@thefnmark\,}#1}
\makeatother 
\renewcommand{\figurename}{\small{Fig.}~}
\sectionfont{\large}
\subsectionfont{\normalsize} 

\fancyfoot{}
\fancyfoot[LO,RE]{\vspace{-7pt}\includegraphics[height=9pt]{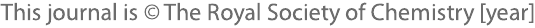}}
\fancyfoot[CO]{\vspace{-7.2pt}\hspace{12.2cm}\includegraphics{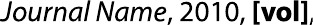}}
\fancyfoot[CE]{\vspace{-7.5pt}\hspace{-13.5cm}\includegraphics{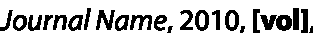}}
\fancyfoot[RO]{\footnotesize{\sffamily{1--\pageref{LastPage} ~\textbar  \hspace{2pt}\thepage}}}
\fancyfoot[LE]{\footnotesize{\sffamily{\thepage~\textbar\hspace{3.45cm} 1--\pageref{LastPage}}}}
\fancyhead{}
\renewcommand{\headrulewidth}{1pt} 
\renewcommand{\footrulewidth}{1pt}
\setlength{\arrayrulewidth}{1pt}
\setlength{\columnsep}{6.5mm}
\setlength\bibsep{1pt}

\newcommand{\onlinecite}[1]{\hspace{-1 ex} \nocite{#1}\citenum{#1}}

\twocolumn[
  \begin{@twocolumnfalse}
\noindent\LARGE{\textbf{
Dynamics of liquid plugs in prewetted capillary tubes: from acceleration and rupture to deceleration and airway obstruction.}}
\vspace{0.6cm}

\noindent\large{\textbf{J.C. Magniez,\textit{$^{a}$} M. Baudoin,\textit{$^{\ast a}$} C. Liu,\textit{$^{a}$} and F. Zoueshtiagh
\textit{$^{a}$}}}\vspace{0.5cm}

\noindent\textit{\small{\textbf{Received Xth XXXXXXXXXX 20XX, Accepted Xth XXXXXXXXX 20XX\newline
First published on the web Xth XXXXXXXXXX 200X}}}

\noindent \textbf{\small{DOI: 10.1039/b000000x}}
\vspace{0.6cm}

  \noindent \normalsize{The dynamics of individual liquid plugs pushed at constant pressure head inside prewetted cylindrical capillary tubes is investigated experimentally and theoretically. It is shown that, depending on the thickness of the prewetting film and the magnitude of the pressure head, the plugs can either experience a continuous acceleration leading to a dramatic decrease of their size and eventually their rupture or conversely, a progressive deceleration associated with their growth and an exacerbation of the airway obstruction. These behaviors are quantitatively reproduced with a simple nonlinear model [Baudoin \textit{et al., Proc. Nat. Ac. Sci. USA}, 2013, \textbf{110}, 859] adapted here for cylindrical channels. Furthermore, an analytical criterion for the transition between these two regimes is derived and successfully compared with extensive experimental data. The potential implications of this work for pulmonary obstructive diseases are discussed.} 
\vspace{0.5cm}
 \end{@twocolumnfalse}
  ]

\section{Introduction}

\footnotetext{\textit{$^{a}$~IEMN, International Laboratory LEMAC/LICS, UMR CNRS 8520, Universit\'{e} de Lille, Avenue Poincar\'e, 59652 Villeneuve d'Ascq, France}}

\footnotetext{$\ast$ Corresponding author. E-mail: michael.baudoin@univ-lille1.fr}

The dynamics of liquid plugs (also called bridges or slugs) is involved in a variety of natural and engineered systems including oil extraction \cite{abb_havre_2000,mp_dimeglio_2011}, flows in porous media \cite{spej_hirasaki_1985}, microsystems, or flows in pulmonary airways. In microsystems, liquid plugs can be used as microreactors \cite{song2003microfluidic,lc_gunther_2004,ace_song_2006,gunther2006multiphase,cep_assmann_2011,mn_ladosz_2016}, since they ensure efficient mixing with no dispersion and controlled diffusion at the interface.  In pulmonary airways, mucus plugs may either form \cite{heil2008mechanics,kamm1989airway,duclaux2006effects,dietze2015films,foroughi2012immiscible,gauglitz1988extended,cassidy1999surfactant} naturally in distal airways of healthy subjects at full expiration \cite{jap_burger_1968,jap_hugues_1970}, or due to diseases leading to an increase of the amount of mucus in the airways' lining, such as cystic fibrosis, chronic obstructive pulmonary diseases (COPD) or asthma \cite{rc_rogers_2007,griese1997pulmonary,wright2000altered,hohlfeld2002role,pof_grotberg_2011}. In this case, liquid plugs dramatically alter the distribution of air inside the airways provoking severe breathing difficulties. Conversely liquid plugs can be injected in the airways for therapy in prematurely born infants to compensate for the lack of surfactants \cite{engle2008surfactant,barber2010respiratory} and thus improve ventilation \cite{espinosa1998meniscus,stevens2007surfactant,engle2008surfactant,halpern2008liquid}, or for drug delivery \cite{acp_vantveen_1998,cepp_nimmo_2002}.

In all the aforementioned examples, a thorough understanding of the dynamics of liquid plugs in channels, and especially their stability to breaking is critical. Indeed, microfluidic microreactors require stable liquid plugs, whereas in the case of pulmonary diseases, the therapy targets plug ruptures to reopen obstructed airways. As a consequence, many experimental \cite{aussillous2000quick,bico2001falling,ijmf_fries_2008,zheng2009liquid,song2011air,baudoin2013airway,hu2015microfluidic}, numerical \cite{pof_fujioka_2005,campana2007stability,fujioka2008unsteady,ubal2008stability,zheng2009liquid,hassan2011adaptive,song2011air} and theoretical \cite{pof_ratulowski_1989,jcis_jensen_2000,a_kreutzer_2005,mn_warnier_2010,baudoin2013airway} studies have investigated the dynamics \cite{aussillous2000quick,bico2001falling,pof_fujioka_2005,campana2007stability,ijmf_fries_2008,ubal2008stability,fujioka2008unsteady,zheng2009liquid,mn_warnier_2010,song2011air,baudoin2013airway} and the rupture \cite{hassan2011adaptive,baudoin2013airway,hu2015microfluidic} of single \cite{aussillous2000quick,bico2001falling,pof_fujioka_2005,campana2007stability,ubal2008stability,fujioka2008unsteady,zheng2009liquid,song2011air,hassan2011adaptive,baudoin2013airway,hu2015microfluidic} or multiple liquid plugs \cite{pof_ratulowski_1989,a_kreutzer_2005,ijmf_fries_2008,mn_warnier_2010,baudoin2013airway} separated by air bubbles in rigid \cite{pof_ratulowski_1989,aussillous2000quick,bico2001falling,a_kreutzer_2005,pof_fujioka_2005,ijmf_fries_2008,fujioka2008unsteady,mn_warnier_2010,song2011air,hassan2011adaptive,baudoin2013airway,hu2015microfluidic} or compliant channels \cite{zheng2009liquid}. The early studies on liquid plugs dynamics were mostly devoted to the fundamental understanding of simple geometric configurations. More recently, complex geometries, such as binary trees reminiscent of the lung structure, have been investigated\cite{song2011air,baudoin2013airway,filoche2015three,stewart2015patterns} due to their relevance for biomedical applications \cite{weibel1984pathway,west1997general,mauroy2004optimal,pedley1977pulmonary,west1986beyond}.
 
In this paper,  the dynamics of liquid plugs in prewetted capillary tubes is studied experimentally and theoretically. This configuration plays a fundamental role in biological systems such as lungs, whose walls are prewetted by mucus, but also in porous media or microsystems when a set of liquid plugs move collectively. It is shown that, depending on the pressure head imposed at the channel entrance and the thickness of the prewetting film, the plugs may either experience a continuous acceleration leading to their rupture or conversely a progressive deceleration leading to their growth. While the former behavior has been observed experimentally \cite{baudoin2013airway}, the latter has, to the best of our knowledge, only been reported in numerical simulations for some specific parameters \cite{fujioka2008unsteady,hassan2011adaptive}. Here, this behavior is evidenced experimentally and rationalized with a model inspired from a previous model developed by Baudoin et al. \cite{baudoin2013airway} adapted to cylindrical channels. This model, which does not contain any fitting parameter, not only enables to quantitatively reproduce the accelerating or decelerating behavior of the liquid plug but also to derive an analytical criterion for the critical pressure at which the transition between the two behaviors occurs.

The first section describes the experimental protocol. The second and third sections are respectively devoted to the dynamics of liquid plugs pushed at constant flow rate and constant pressure head. For each case, an analytical model is developed and compared to experiments. Finally the last section discusses the relevance of these results for pulmonary obstructive diseases.


\section{Methods}

\subsection{Experimental protocol}

\begin{figure}[htbp]
\centering
  \includegraphics[width=9cm]{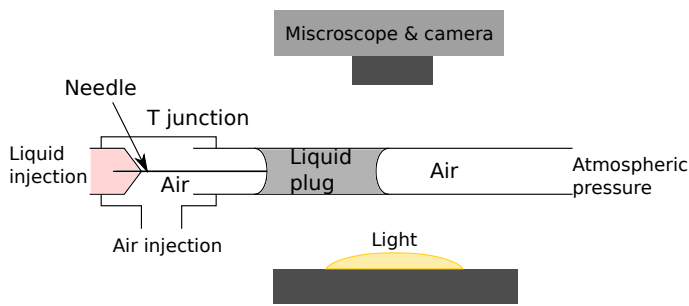}
  \caption{Sketch of the experimental setup}
  \label{fig:dispositif}
\end{figure}

In the following, we investigate the dynamics of a liquid plug pushed at constant flow rate or constant pressure head inside a cylindrical prewetted rigid glass capillary tube (Ringcaps 20/40 $\mu$L) of inner radius $R = $500 $\mu$m (see Fig. \ref{fig:dispositif}). The tube prewetting is achieved with a thorough protocol: \textit{First}, the glass tubes are cleaned successively with acetone, isopropanal, dichloromethane and piranha solutions (a mixture of sulfuric acid ($\mathrm{H_2SO_4}$) and hydrogen peroxide ($\mathrm{H_2O_2}$)) to achieve perfectly wetting surfaces.  \textit{Second}, a plug is injected inside the tube though a T-junction by pushing at constant flow rate a prescribed amount of liquid with a syringe pump (see Fig. \ref{fig:dispositif}). This plug is then pushed with a second syringe pump at a constant air flow rate (and thus constant velocity)  to create a prewetting film of constant thickness $h_d$ on the tube walls.

\begin{table}[htbp]
\small
  \caption{Mechanical proprieties of the liquids used in experiments at 20$^\circ$C\cite{glycerine1963physical}}
  \begin{tabular*}{0.5\textwidth}{@{\extracolsep{\fill}}llll}
    \hline
    Liquid & $\sigma$ (mN/m) & $\mu_l$ (mPa.s)& $\rho_l$ (kg/m$^3$)\\
    \hline
    Perfluorodecalin & 19.3 & 5.1 & 10$^3$\\
    Aqueous glycerol 10\% & 71 & 1.31 & 1.03.10$^3$\\
    Aqueous glycerol 60\% & 67 & 10.8 & 1.15.10$^3$\\
    Aqueous glycerol 85\% & 65 & 223 & 1.21.10$^3$\\
    \hline
  \end{tabular*}
  \label{table:carac}
\end{table}

Then, a second liquid plug is injected inside the prewetted capillary tube with the same protocol and pushed either at constant flow rate with a syringe pump or at constant pressure with a Fluigent MFCS pressure controller. Four different liquids (perfluorodecalin and three aqueous glycerol mixtures of respectively $10\%$, $60\%$ and $85\%$ mass concentration of glycerol) have been used to explore different regimes. The properties of these liquids are summarized in Table \ref{table:carac}.  

\begin{figure}[htbp]
  \centering
  \includegraphics[width=9cm]{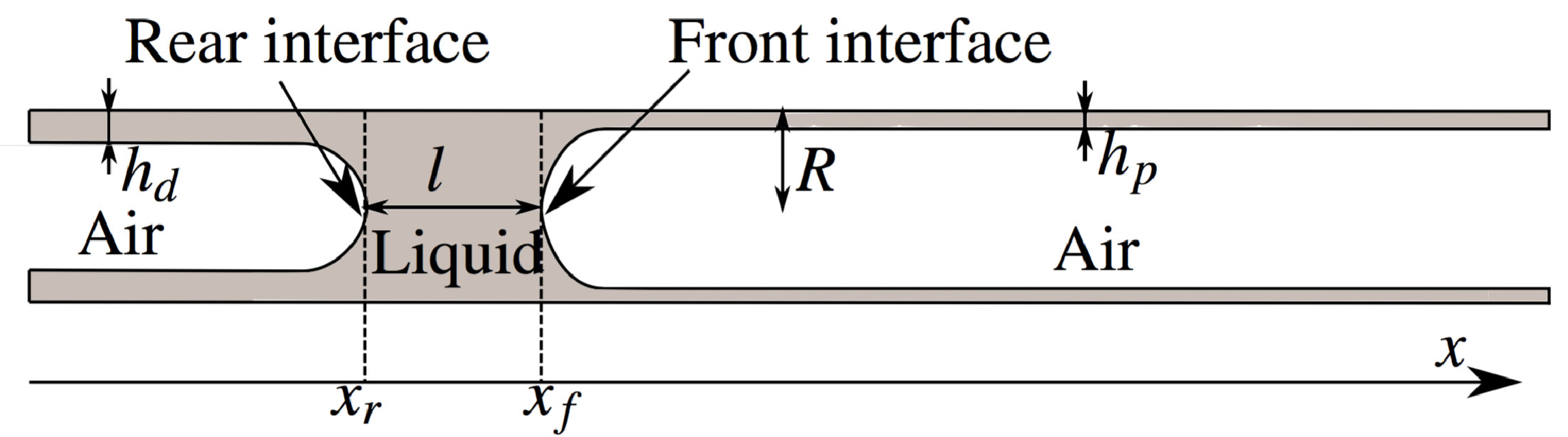}
  \caption{Sketch of the problem considered in this paper}
  \label{fig:problem}
\end{figure}

The plugs dynamics are characterized by the evolution of its speed $U = d x_r / dt$  and its size $l= x_f - x_r$, with $x_r$ and $x_f$ the positions of the rear and front interface respectively (see Fig. \ref{fig:problem}). It is recorded with a Photron SA3 high speed camera  mounted on a Z16 Leica Macroscope. To avoid spurious diffraction of light by the cylindrical walls, the capillary tube is immersed in an index-matching liquid. Then the images are post-processed with ImageJ software. Since the viscosity and surface tension of aqueous glycerol mixtures are sensitive to temperature, the liquid temperature is measured for each experiment and the data provided in ref.\onlinecite{glycerine1963physical} are used to interpolate the liquid properties at the appropriate temperature.

\subsection{Dimensional analysis of the flow regime}

\label{ss:dimensional}


In these experiments, the capillary number $Ca = \mu_l U / \sigma$, the Bond number $Bo = \rho_l g R^2 / \sigma$ and the Weber number $We =\rho U^2 R / \sigma$ remain small ($1.29 \times 10^{-4}<Ca<5 \times 10^{-2}$, $3.5  \times 10^{-2}<Bo< 0.12$, and $1.2 \times 10^{-4} <We< 0.27$  ), with $\mu_l$ the liquid viscosity, $\sigma$ the surface tension, $\rho_l$ the density of the liquid and $g$ the gravitational acceleration. These three numbers compare respectively viscous effects, gravity and inertial effects to surface tension, indicating that surface tension is globally dominant.

Nevertheless, while viscous effects can be neglected in the central part of the meniscus called the "static meniscus", they still play an important role near the walls, where the thin prewetting and trailing films lead to high velocity gradients and thus viscous stresses. 

\begin{table}[htbp]
\small
  \caption{Comparison  of the range of variation of the capillary number $Ca= \mu_l U / \sigma$ measured experimentally with the  critical capillary number $Ca_c$  introduced by Aussilous \& Qu\'{e}r\'{e} \cite{aussillous2000quick} above which inertial effects become significant .}

 \begin{tabular*}{0.5\textwidth}{@{\extracolsep{\fill}}lcc}
    \hline
    Liquid & $Ca$ & $Ca_c$  \\
    \hline
    Perfluorodecalin & 1.29 $\times$ 10$^{-4}<Ca<$4.8 $\times$ 10$^{-2}$ & 3.6 $\times$ 10$^{-1}$  \\
    Aqueous glycerol 10\% & 1.35 $\times$10$^{-4}<Ca<$1.6 $\times$ 10$^{-3}$ & 1.7 $\times$ 10$^{-2}$   \\
    Aqueous glycerol 60\% & 1.29 $\times$ 10$^{-4}<Ca<$5.0 $\times$ 10$^{-2}$ & 3.9 $\times$ 10$^{-1}$  \\
    Aqueous glycerol 85\% & 4.10 $\times$ 10$^{-4}<Ca<$3.1 $\times$ 10$^{-2}$ & 36.1   \\
    \hline
  \end{tabular*}
 
  \label{table:sdim}
\end{table}

Concerning inertial effects, the weak to moderate Reynolds numbers $Re = \rho_l U R / \mu_l$ in the experiments (3.10$^{-4}$ $<Re<$ 5.58) indicate that viscous effects are generally dominant over inertial effects. To quantify more precisely their impact, we measured the capillary number in our experiments and compared it to the critical capillary number $Ca_c$ introduced by Aussilous and Qu\'{e}r\'{e} \cite{aussillous2000quick} (see table \ref{table:sdim}), above which inertial effects are expected to play a significant role. This critical number was derived from dimensional analysis by these authors and validated experimentally. This comparison shows that inertial effect can be safely neglected since $Ca$ always lies significantly below $Ca_c$. In addition, the influence of inertial effects on the pressure drop induced by bubbles in capillaries was studied numerically by Kreutzer et al. \cite{a_kreutzer_2005} and shown to play a negligible role for $Re<10$  at capillary numbers of the same order of magnitude as the one measured in the present study.

\begin{table}[htbp]
\small
  \caption{Comparison of the duration of the experiments $\Delta t$ to the characteristic time associated with unsteady viscous effects $\tau_v = R^2 / \rho_l \mu_l$ for each fluid.}
 \textcolor{black}{
  \begin{tabular*}{0.5\textwidth}{@{\extracolsep{\fill}}lcc}
    \hline
    Liquid  & $\Delta t$ (ms) & $\tau_v$ (ms)  \\
    \hline
    Perfluorodecalin & 1200$<\Delta t<$1300 & 50  \\
    Aqueous glycerol 10\% & 800<$\Delta t$<4800 & 200  \\
    Aqueous glycerol 60\% & 900<$\Delta t$<1400 & 3  \\
    Aqueous glycerol 85\% & 2300<$\Delta t$<22000& 1.5  \\
    \hline
  \end{tabular*}
  }
  \label{table:sdim2}
\end{table}

Finally, the role of unsteady viscous effect can be quantified by comparing the duration of the experiments $\Delta t$ to the characteristic time $\tau_v = R^2 / \rho_l \mu_l$ associated with unsteady viscous effects. The duration of the experiments is substantially lower than $\tau_v$ for all experiments (see table \ref{table:sdim2}). Thus unsteady viscous effects are expected to play a negligible role. 

As a consequence, the liquid plugs dynamics in the present experiments can be analyzed as a quasi-static flow mainly determined by viscous and capillary effects
 
\subsection{Prewetting film thickness}

\begin{figure}[htbp]
\centering
  \includegraphics[height=5cm]{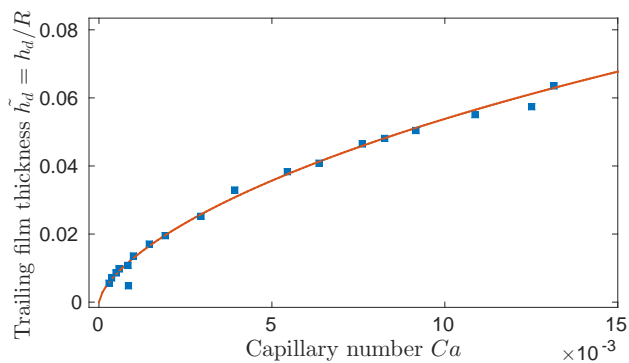}
  \caption{Thickness $\tilde{h_d} = h_d/R$ of the film deposited by the motion of a liquid plug at constant velocity $Ca = \mu_l U / \sigma$ in a perfectly wettable cylindrical channels of radius $R$. The blue points correspond to experimental measurements and the red line to Aussillous law obtained from dimensional analysis \cite{aussillous2000quick}.}
  \label{fig:aussillous}
\end{figure}

A key parameter for the realization of these experiments is the deposition of a prewetting film of controlled thickness on the walls. This film is deposited by pushing a first liquid plug at constant flow rate inside the tube. Indeed, it is well known since the seminal work of Bretherton \cite{bretherton1961motion} that the motion of a perfectly wetting liquid plug at constant speed inside a cylindrical  channel leaves a liquid trailing film of constant thickness in its wake. Bretherton's calculations (only valid at very low capillary numbers $Ca < 10^{-3}$) was further extended to higher capillary numbers by Aussillous and Qu\'{e}r\'{e},\cite{aussillous2000quick} through  the combination of dimensional analysis and experiments to determine the missing constants:
\begin{equation}
  \frac{h_d}{R} = \frac{1.34 \, Ca^{2/3}}{1+2.5\times 1.34\, Ca^{2/3}}
\label{eq:thickness}
\end{equation}
We verified this formula by monitoring the decrease of the prewetting plug size resulting from the deposition of a trailing liquid film. Excellent agreement was achieved as seen on Fig. \ref{fig:aussillous}, thus underlining the accuracy of our "prewetting protocol".

\section{Results}

\subsection{Plug dynamics at constant flow rate}

\begin{figure}
  \centering
   \includegraphics[width=9cm]{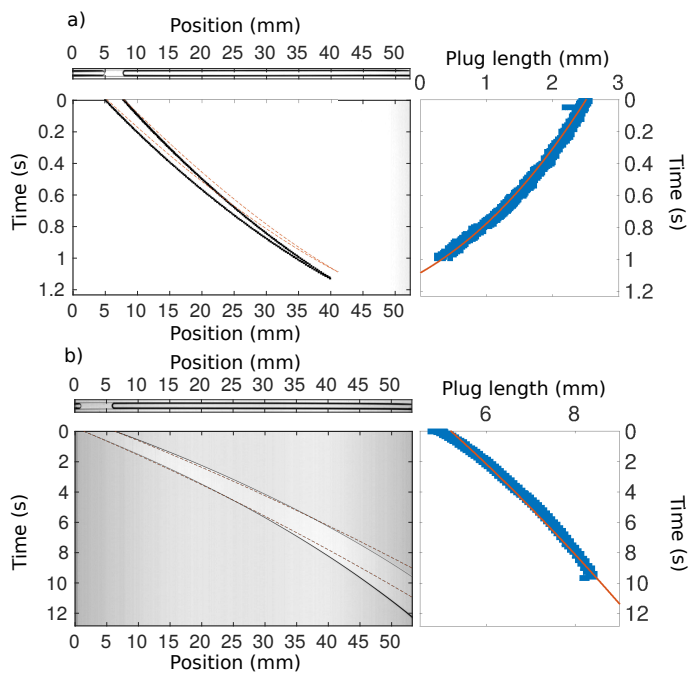}
  \caption{Dynamics of a perfluorodecalin liquid plug pushed at constant flow rate with a syringe pump.
a) Plug size decrease obtained for a prewetting film thinner than the trailing film: $h_p/R=0.038$ and $h_d/R=0.058$, b) Plug size increase obtained for a prewetting film thicker than the trailing film: $h_p/R=0.042$ and $h_d/R=0.016$. \textit{Left:} Spatiotemporal diagram displaying the gray values along the center line of the channel as a function of time. The two black lines correspond to the positions of the rear and the front interfaces. The slope of the left black line gives the plugs speed $U(t)$ and the horizontal distance between the left and the right black line, the size of the plug $l(t)$. \textit{Right:} Evolution of the volume of the liquid plug. Blue points: experimental values. Red line: theoretical curve obtained from equation (\ref{eq:volume_evolution}).}.
  \label{fig:debit}
\end{figure}

When a liquid plug is pushed  inside a \textit{prewetted} capillary tube at constant flow rate $Q$, it leaves a film of constant thickness $h_d$ in its wake and recovers some liquid from the prewetting film of thickness $h_p$. When $h_d > h_p$, the plug loses more liquid  than it gains from the prewetting film. Since this balance remains constant over time, the plug size decreases linearly as illustrated on Fig. \ref{fig:debit}a. Conversely, when $h_p> h_d$, its size increases linearly (see Fig. \ref{fig:debit}b).

A balance of the lost and gained liquid gives:
\begin{equation}
dV=-dx_r((R^2-(R-h_d)^2)\pi+dx_f\pi((R^2-(R-h_p)^2) \nonumber
\end{equation}  
with $V$ the volume of the plug, $dx_r=Udt$ and $dx_f=dx_r\frac{(R-h_d)^2}{(R-h_p)^2}$ by volume conservation. An equation for the plug length is then simply obtained from the relation $dl = dV / \pi R^2$:
\begin{equation}
\frac{dl}{dt}=U \left[ 1-\frac{(R-h_p)^2}{(R-h_d)^2} \right]
\label{eq:volume_evolution}
\end{equation}
with $h_d$ and $h_p$ related to the first and second plug velocity $U=\frac{Q}{\pi R^2}$ according to equation (\ref{eq:thickness}). The predictions given by this equation are compared to the experimentally measured evolution of the liquid plug length (see red curves on Fig. \ref{fig:debit}) leading to an excellent quantitative agreement.

\subsection{Plug dynamics at constant pressure}

\label{ss:pressure}

\begin{figure}
  \centering
   \includegraphics[width=9cm]{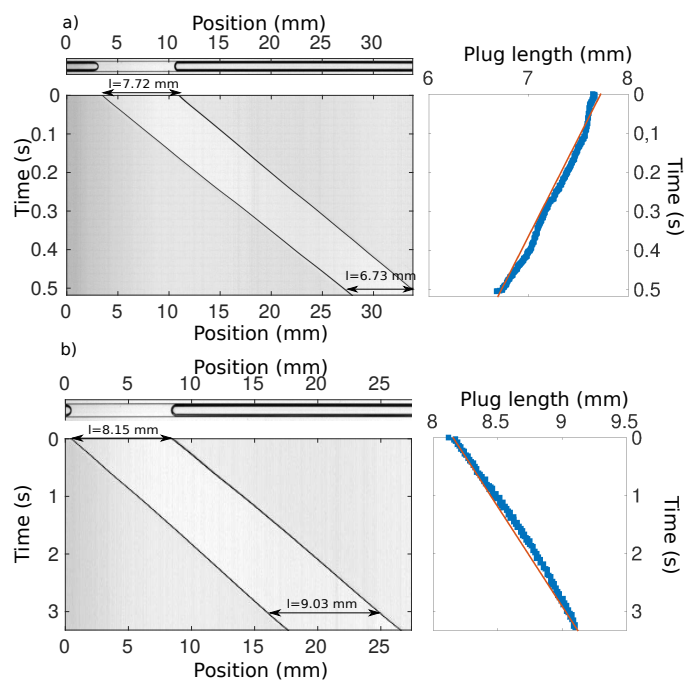}
  \caption{Dynamics of a perfluorodecalin liquid plug pushed at constant pressure head with the MFCS pressure controller. a) Plug acceleration observed for $h_p/R=0.012$ and $\Delta P= 0.3$ mbar, b) Plug deceleration observed for $h_p/R=0.0482$ and $\Delta P= 0.1$ mbar. \textit{Left:} Spatiotemporal diagram displaying the gray values along the center line of the channel as a function of time. The two black solid lines correspond respectively to the position of the rear and front interfaces. The slope of the left black solid line gives the plugs speed $U(t)$ and the horizontal distance between the left and the right black line, the size of the plug $l(t)$. The red dotted lines correspond to predictions obtained from equations  (\ref{eq:evolution_dynamic}), (\ref{eq:ltilde}), (\ref{eq:thethadca}) and (\ref{eq:thickness}). \textit{Right:} Evolution of the length of the liquid plug. Blue points: experimental values. Red continuous line: theoretical predictions.}
  \label{fig:P_relie}
\end{figure}

The liquid plugs dynamics at constant pressure head is more complex since the plug velocity $U$ is no more constant and strongly depends on the plug size $l$. Two regimes can be distinguished:  (i) the liquid plug accelerates continuously and its size decreases until its rupture  (Fig. \ref{fig:P_relie}a), (ii) the liquid plug decelerates with a progressive increase of its size (Fig. \ref{fig:P_relie}b)

These regimes can be theoretically derived by adapting the rectangular channel model of ref. \citenum{baudoin2013airway} to cylindrical channels. Assuming air pressure loss negligible compared to the one induced by the liquid plug, the total pressure drop $\Delta P$ across the channel can be considered as the sum of the pressure drops $\Delta P^{int}_{rear}$ and $\Delta P^{int}_{front}$ at the rear and front interfaces and the viscous pressure drop in the bulk of the plug $\Delta P^{bulk}_{visc}$:
\begin{equation}
\Delta P=\Delta P^{int}_{rear}+\Delta P^{int}_{front}+\Delta P^{bulk}_{visc}.
\label{eq1}
\end{equation}
$\Delta P^{int}_{rear}$  can be inferred from Bretherton's model \cite{bretherton1961motion}:
\begin{equation}
\Delta P^{int}_{rear}= \frac{2\sigma}{R} \left( 1+1.79(3Ca)^{2/3} \right).
\label{eq2}
\end{equation}
$\Delta P^{int}_{front}$ was calculated by Chebby\cite{chebbi2003deformation} for a liquid plug moving in a prewetted cylindrical tube:
\begin{equation}
\Delta P^{int}_{front}= -\frac{2\sigma}{R}\cos(\theta_d),
\label{eq3}
\end{equation}
with $\theta_d$ the dynamic apparent contact angle of the front interface, whose value is related to $Ca$ according to:
\begin{equation}
\tan(\theta_d)=3^{1/3}Ca^{1/3}F(3^{-2/3}Ca^{-2/3} \frac{h_p}{R} \cos(\theta_d))
\label{eq:evolution_theta}
\end{equation} 
with $F(x)=\sum\limits_{n=0}^3 \, b_n \, log_{10}(x)^n$, $b_0=$1.4, $b_1=$-0.59, $b_2=-3.2 \times 10^{-2}$ and $b_3=3.1 \times 10^{-3}$.

Finally $\Delta P^{bulk}_{visc}$ can be approximated from Poiseuille's law:
\begin{equation}
\Delta P^{bulk}_{visc}=\frac{8\mu lU}{R^2}
\label{eq4}
\end{equation}

Dimensionless form of these equations is obtained by introducing the characteristic length scale $R$, pressure $2 \sigma / R$, time scale $\mu R / \sigma$ and  speed $\sigma/\mu$. By combining equations (\ref{eq1}), (\ref{eq2}), (\ref{eq3}) and (\ref{eq4}), one obtains:
\begin{equation}
\Delta  \tilde{P}=1+1.79(3Ca)^{2/3}-\cos(\theta_d)+4\tilde{l}Ca
\label{eq:evolution_dynamic}
\end{equation}
where the index $\sim$  indicates dimensionless functions and $\theta_d$ is related to $Ca$ according to equation (\ref{eq:evolution_theta}). Since $\Delta \tilde{P}$ is constant, this equation enables to compute the dimensionless velocity (i.e., the capillary number) as a function of the dimensionless plug size $\tilde{l}$. The latter is computed from the dimensionless form of equation (\ref{eq:volume_evolution}):
\begin{equation}
\frac{d\tilde{l}}{d\tilde{t}}=\frac{Ca}{\pi} \left[ 1-\frac{(1-\tilde{h_p})^2}{(1-\tilde{h_d})^2} \right]
\label{eq:ltilde}
\end{equation}
with $\tilde{h_d} = h_d / R$ given by equation (\ref{eq:thickness}). Thus, equations (\ref{eq:evolution_dynamic}), (\ref{eq:ltilde}), (\ref{eq:evolution_theta}) and (\ref{eq:thickness}) form a closed set of equations. Equation (\ref{eq:evolution_dynamic}) can be further simplified by considering the low $Ca$ (and thus $\theta_d$) limit \cite{chebbi2003deformation} : $\cos(\theta_d)\approx 1-\frac{\theta_d^2}{2}$. At the lowest order the equation (\ref{eq:evolution_theta}) reduces to a quadratic equation whose solution is:
\begin{equation}
\theta_d=\frac{-1+\sqrt{1+4CD}}{2C}
\label{eq:thethadca}
\end{equation}
with $A=(3Ca)^{-2/3}\tilde{h_p}$, $B=(3Ca)^{1/3}$, $C=(\frac{b_1}{\log(10)}+\frac{b_2\log_{10}(A)}{\log(10)}+\frac{b_33\log_{10}(A)^2}{2\log(10)})B$ and $D=(b_0+b_1\log_{10}(A)+b_2\log_{10}(A)^2+b_3\log_{10}(A)^3)B$

Equations (\ref{eq:evolution_dynamic}), (\ref{eq:ltilde}), (\ref{eq:thethadca}) and (\ref{eq:thickness}) are solved by using an Euler method for the discretization of the nonlinear differential equation (\ref{eq:ltilde}), coupled with a dichotomy method to solve equation (\ref{eq:evolution_dynamic}) at each time step. This model reproduces quantitatively both the catastrophic acceleration and deceleration regimes associated with a plug size decrease and increase respectively (see red lines on Fig. \ref{fig:P_relie}).

\begin{figure}
  \centering
   \includegraphics[width=9cm]{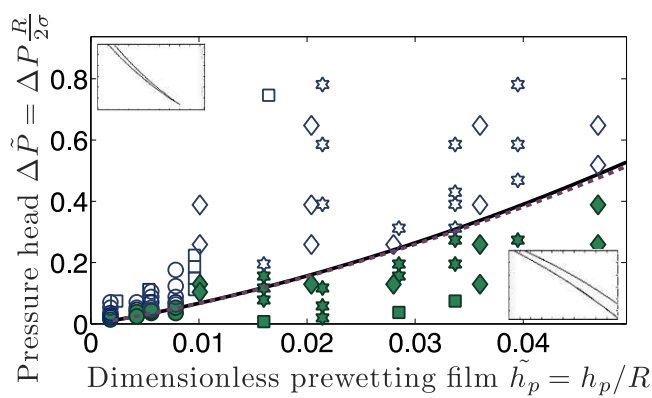}
  \caption{Phase diagram of the dynamics of a 3 $\mu$L (length $l=$3.80 mm) liquid plug  pushed at different pressure head $\Delta \tilde{P}$ inside tubes covered with prewetting films of different thicknesses $\tilde{h}_p$. Blue empty points correspond to the acceleration regime and green filled points to the deceleration regime. The black continuous line is obtained from the the analytical expression derived in this paper. The purple dashed line is obtained from the theory without the low $\theta_D$ approximation. Star : aqueous glycerol 85$\%$, Square : aqueous glycerol 60 $\%$, Circle : aqueous glycerol 10 $\%$, Diamonds : perfluorodecalin}
  \label{fig:transition_P}
\end{figure}

\begin{figure}
  \centering
   \includegraphics[width=9cm]{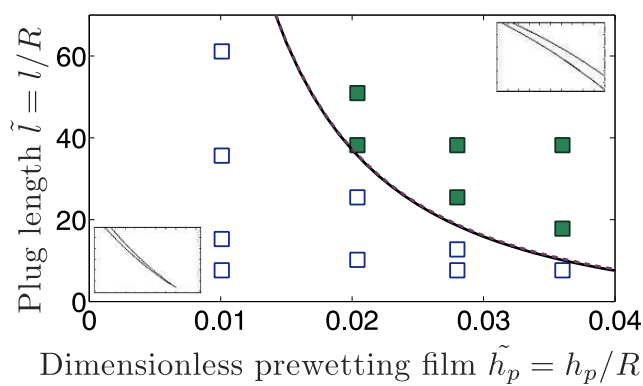}
  \caption{Phase diagram of the dynamics of liquid plug of different initial lengths $\tilde{l}$ pushed at a prescribed pressure head $\Delta P = 0.3$ mbar inside tubes covered with a prewetting films of different thicknesses $\tilde{h}_p$. Orange empty points correspond to the acceleration regime and green filled points to the deceleration regime. The black continuous line is obtained from the analytical expression derived in this paper. The purple dashed line is obtained from the theory without the low $\theta_D$ approximation.}
  \label{fig:transition_L}
\end{figure}

This model further provides an analytical expression for the critical pressure at which the transition between the two regimes occurs, i.e. when $\tilde{h}_d=\tilde{h}_p$ ($d \tilde{l} / d \tilde{t} = 0$). From eq. (\ref{eq:thickness}), one can compute the critical initial capillary number at the transition:
\begin{equation}
Ca_c=\left(\frac{\tilde{h_p}}{1.34(1-2.5\tilde{h_p})}\right)^{3/2}
\end{equation}
Insertion of this expression into equation (\ref{eq:evolution_dynamic}) provides the transition critical pressure head $\Delta \tilde{P}_c$ as a function of the prewetting film thickness $\tilde{h}_p$ and the initial plug length $\tilde{l}$. This criterion was verified experimentally (see Fig. \ref{fig:transition_P}) over 70 experiments performed for different pressure heads, liquids and prewetting film thicknesses and constant initial plug length $\tilde{l}$ . Excellent agreement is achieved between the derived formula and experiments. This transition was also studied experimentally at constant pressure head $\Delta \tilde{P}$ for various initial plug length $\tilde{l}$ (see Fig.\ref{fig:transition_L}). Again, excellent agreement is achieved with the analytical formula provided in this paper. The sources of experimental uncertainties are (i) the response time of the MFCS controller (about  $100$ ms ), (ii) the uncertainty of 7.5 $\mu$bar on the pressure imposed by the MFCS controller and (iii) the variations of the initial plug length $l$ (about 5$\%$).

The present theory is also in qualitative agreement with the numerical simulations of Hassan et al. \cite{hassan2011adaptive}, who found a transition between acceleration and deceleration dynamics for $\tilde{h}_p= 0.09-0.10$ and $\tilde{l}=1$ at $\Delta \tilde{P}_c= 0.5$, and for $\tilde{h}_p= 0.05$, $\tilde{l}=1$ at $\Delta \tilde{P}_c = 0.22$. The present model gives respectively $\Delta \tilde{P}_c = 0.65-0.75$ and $\Delta \tilde{P}_c= 0.3$. The discrepancy between the values predicted by our model and those of Hassan et al. may come from the systematic initial plug growth observed in their simulations, independently of the regime,  which is not observed experimentally.

\section{Discussion}

\subsection{Critical assessment of the models validity.}

The models derived in this paper rely on the assumption of low capillary, Bond, Weber and Reynolds numbers. This hypothesis is well verified as discussed in section \ref{ss:dimensional}. Furthermore, since the duration of the experiments is significantly larger than the characteristic time associated with unsteady viscous effects, quasi-static models can be used even if the plug velocity evolve as a function of time. These approximations authorize the use of simple analytic expressions of the rear, front and bulk pressure drop and of the thickness of the trailing film deposited on the wall.

In section \ref{ss:pressure}, we made the additional approximation that the dynamic contact angle $\theta_D$ is small. It enables the derivation of a simple analytical expression of the critical pressure at which the transition between acceleration and deceleration dynamics occurs. This approximation is nevertheless not necessary to determine the solution of the problem, which can be solved numerically. To quantify the impact of this approximation, we calculated the critical pressure head without this simplification. The results are shown on Fig. \ref{fig:transition_P} and \ref{fig:transition_L} (purple dashed line). The comparison with the analytical expression shows no significant difference assessing the validity of this simplification.

Finally, the expressions used in this paper for the trailing film thickness and pressure drop rely on the assumption that the plug length is significantly larger than the tube radius. The effect of limited plugs length on the trailing film thickness was studied numerically by Campana et al \cite{campana2007stability}. Their result indicate a weak influence of the plug length on the thickness of the trailing film ($<2 \%$) at Reynolds number $Re<10$ and only for plugs length $\tilde{l} < 2$. We found nevertheless no study quantifying the effect of limited plugs length on the pressure drop.

All these elements confirm the validity of our assumptions and explain the quantitative agreement of this simplified model with the experiments. Nevertheless some discrepancies remain (see Fig. \ref{fig:debit} and \ref{fig:P_relie}). For the flow rate driven experiments,  we found that these differences are mainly due to some fluctuations (a few percent) in the flow rate imposed by the syringe pump.  For the longest experiments (deceleration regime), another envisioned source of discrepancy between the model and the experiments might be the attraction of the prewetting liquid film by the plug. This might explain the slightly larger plug growth at the beginning of the experiment and slightly smaller at the end.

\subsection{Relevance of this study for pulmonary obstructive diseases} 

The primary purpose of the lung is to promote gas exchanges ($\mathrm{CO_2}$ / $\mathrm{O_2}$) between the bloodstream and the airways. These exchanges mainly occur in the alveoli, highly fragile organs with extremely thin walls. To prevent alveoli occlusion, the air flowing in the lung is purified by the so-called mucociliary apparatus, a thin mucus layer produced all along the airway tree and conveyed by micrometric vibrating cilia covering the airways. The mucus captures the impurities present in the airways lumen and convey them to the throat where they are swallowed. In healthy subjects, the thickness of this mucus layer is typically $5$ to $10$ $\mu$m \cite{pedley1977pulmonary,widdicombe2002regulation}. However, some diseases like chronic obstructive pulmonary disease \cite{pedley1977pulmonary,King}, cystic fibrosis \cite{widdicombe2002regulation} or asthma can dramatically increase the amount of mucus present in the airways (thickness of the mucus film up to $25 \%$ of the airways inner radius), leading to the formation of liquid plugs through a Rayleigh-Plateau instability \cite{heil2008mechanics}. These plugs considerably alter the distribution of air inside the bronchial tree. Reopening of obstructed airways may occur either during normal breathing cycle or coughing, due to the accelerative dynamics of the plugs  \cite{hassan2011adaptive,baudoin2013airway,hu2015microfluidic}. Nevertheless, such accelerative cascade can only occur above the critical pressure threshold described in this paper. Otherwise, the plugs' motion will only result in an amplification of the occlusion.

\begin{figure}
  \centering
   \includegraphics[width=0.5\textwidth]{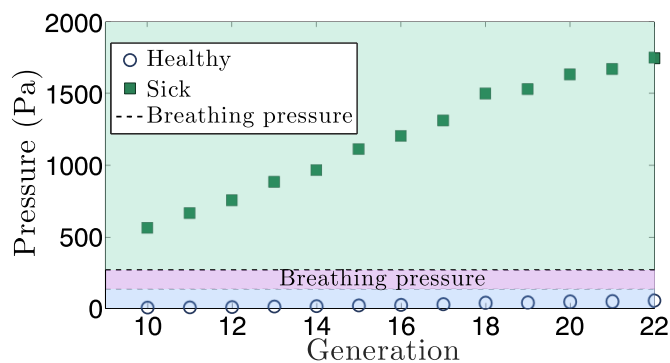}
  \caption{Critical pressure for the transition between acceleration and deceleration regimes of mucus plugs dynamics inside pulmonary airways for different generations. The critical pressure is calculated from the formula presented in this paper and physiological data available in the literature. Blue circle points: healthy subject (mucus larger thickness of 10 $\mu$m). Green rectangular points: sick patient (mucus layer of 25$\%$ of the radius of the airways). The purple (central) area represents the normal breathing pressure. Points lying in the blue region (lower region) will lead to an acceleration of the liquid plugs and thus reopening of airways obstructed by liquid plugs. While points in the green region will lead to the deceleration of the liquid plug and thus increase of the plug size. Color online}
  \label{fig:physio}
\end{figure} 

In this section, we provide a coarse estimate of which regime (acceleration or deceleration) is expected to be dominant in healthy and sick human lungs based on physiological data available in the literature and on our analytical results. The human lung is a binary tree structure made of about 23 generations with decreasing airways diameters \cite{pedley1977pulmonary,weibel1984pathway,west1986beyond}. Here, we only consider distal airways (generations 9-10 to 23), where the assumptions of our model are valid (Bo$<$1). The size of the plugs formed through Rayleigh-Plateau instability is estimated as $l \sim \lambda(1-(1-h/R)^2)$ with $\lambda=2\pi\sqrt{2}R$ the most unstable wavelength for this instability\cite{duclaux2006effects}. Since, the length of the bronchus \cite{west1986beyond} is comparable to the wavelength $\lambda$ all along the lung tree, only one plug is expected to form for each generation of the lung. Some values of the viscosity and surface tension of the mucus are respectively provided in Grotberg et al.\cite{grotbergpulmonary} and Heil et al\cite{heil2002airway} papers: $\mu \sim 10^{-3}$ kg m$^{-1}$ $s^{-1}$ and $\sigma \sim 20.10^{-3}$ N/m. 

With these data, the critical pressure head $\Delta P_c$ can be estimated for each generation and compared to the breathing pressure magnitude (133 Pa $< \Delta P <$ 266 Pa) in normal breathing conditions for an adult  \cite{marieb2014anatomie} (see Fig. \ref{fig:physio}). If $\Delta P_c$ is smaller than the breathing pressure, the liquid plug should accelerate (and eventually rupture if the breathing cycle is long enough). Conversely if it is higher, the congestion shall worsen during the breathing cycle. Our estimation suggests that for healthy subjects, the liquid plug shall accelerate during the breathing cycle. For sick patients however, the breathing cycle shall result in an increase of the congestion. In this case, reopening could only be achieved by applying stronger pressure (e.g. by coughing) to rupture the plugs. The implications of the critical pressure introduced in this paper on pulmonary obstructive diseases remain nevertheless prospective at this stage. A quantitative investigation should include the non-Newtonian properties of the mucus \cite{King,zamankhan2012steady,hu2015microfluidic}, their evolution with respect to the pathology, the presence of pulmonary surfactants \cite{pof_grotberg_2011}, the role of the  flexibility of the last generations of the lung \cite{zheng2009liquid,stewart2015patterns}, the complexity of the breathing cycle and the pressure distribution in the lung network. The orders of magnitude provided here nervertheless suggest further investigation in this direction.

\section{Conclusion}

Despite extensive studies on the dynamics of liquid plugs in capillaries, the effect of a prewetting film on the plugs stability to breaking has never been considered experimentally nor theoretically. Yet, a liquid film is present in most practical situations, either due to a natural production in the lung tree or due to the film deposited by other plugs in the case of multiple plugs motion in simple channels or complex networks. In this paper, we show both experimentally and theoretically that two distinct regimes can be observed when liquid plugs are pushed at constant pressure head inside prewetted capillary tubes: Either the plug will continuously accelerate leading to a decrease of its size and eventually its rupture or decelerate (and thus grow) leading to a dramatic increase of the flow obstruction.

We provide an analytical criterion for the transition between these two regime, which is successfully validated experimentally. The implication of this work for pulmonary congestions is discussed thoroughly. We show that it might provide some fundamental elements toward the understanding of plugs persistence in sick patients airways.

\section*{Acknowledgment}

It is a pleasure to thank Ilyesse Bihi and Theo Mathurin for fruitful discussions and suggestions.

\footnotesize{
\bibliography{rsc} 

\providecommand*{\mcitethebibliography}{\thebibliography}
\csname @ifundefined\endcsname{endmcitethebibliography}
{\let\endmcitethebibliography\endthebibliography}{}
\begin{mcitethebibliography}{62}
\providecommand*{\natexlab}[1]{#1}
\providecommand*{\mciteSetBstSublistMode}[1]{}
\providecommand*{\mciteSetBstMaxWidthForm}[2]{}
\providecommand*{\mciteBstWouldAddEndPuncttrue}
  {\def\EndOfBibitem{\unskip.}}
\providecommand*{\mciteBstWouldAddEndPunctfalse}
  {\let\EndOfBibitem\relax}
\providecommand*{\mciteSetBstMidEndSepPunct}[3]{}
\providecommand*{\mciteSetBstSublistLabelBeginEnd}[3]{}
\providecommand*{\EndOfBibitem}{}
\mciteSetBstSublistMode{f}
\mciteSetBstMaxWidthForm{subitem}
{(\emph{\alph{mcitesubitemcount}})}
\mciteSetBstSublistLabelBeginEnd{\mcitemaxwidthsubitemform\space}
{\relax}{\relax}

\bibitem[Havre \emph{et~al.}(2000)Havre, Stornes, and Stray]{abb_havre_2000}
J.~Havre, K.~Stornes and H.~Stray, \emph{ABB Review}, 2000, \textbf{4},
  55--63\relax
\mciteBstWouldAddEndPuncttrue
\mciteSetBstMidEndSepPunct{\mcitedefaultmidpunct}
{\mcitedefaultendpunct}{\mcitedefaultseppunct}\relax
\EndOfBibitem
\bibitem[Di~Meglio(2011)]{mp_dimeglio_2011}
F.~Di~Meglio, \emph{PhD thesis}, Ecole Nationale sup\'{e}rieure des mines de
  Paris, 2011\relax
\mciteBstWouldAddEndPuncttrue
\mciteSetBstMidEndSepPunct{\mcitedefaultmidpunct}
{\mcitedefaultendpunct}{\mcitedefaultseppunct}\relax
\EndOfBibitem
\bibitem[Hisaraki and Lawson(1985)]{spej_hirasaki_1985}
G.~Hisaraki and J.~Lawson, \emph{Soc. Petr. Eng. J.}, 1985, \textbf{25},
  176--190\relax
\mciteBstWouldAddEndPuncttrue
\mciteSetBstMidEndSepPunct{\mcitedefaultmidpunct}
{\mcitedefaultendpunct}{\mcitedefaultseppunct}\relax
\EndOfBibitem
\bibitem[Song \emph{et~al.}(2003)Song, Tice, and
  Ismagilov]{song2003microfluidic}
H.~Song, J.~D. Tice and R.~F. Ismagilov, \emph{Angew. Chem-Ger. Edit}, 2003,
  \textbf{115}, 792--796\relax
\mciteBstWouldAddEndPuncttrue
\mciteSetBstMidEndSepPunct{\mcitedefaultmidpunct}
{\mcitedefaultendpunct}{\mcitedefaultseppunct}\relax
\EndOfBibitem
\bibitem[Gunther \emph{et~al.}(2004)Gunther, Khan, Thalmann, Trachsel, and
  Jensen]{lc_gunther_2004}
A.~Gunther, S.~Khan, M.~Thalmann, F.~Trachsel and F.~Jensen, \emph{Lab Chip},
  2004, \textbf{4}, 278--286\relax
\mciteBstWouldAddEndPuncttrue
\mciteSetBstMidEndSepPunct{\mcitedefaultmidpunct}
{\mcitedefaultendpunct}{\mcitedefaultseppunct}\relax
\EndOfBibitem
\bibitem[Song \emph{et~al.}(2006)Song, Chen, and Ismagilov]{ace_song_2006}
H.~Song, D.~Chen and R.~Ismagilov, \emph{Angew. Chem. Int. Edit.}, 2006,
  \textbf{45}, 7336--7356\relax
\mciteBstWouldAddEndPuncttrue
\mciteSetBstMidEndSepPunct{\mcitedefaultmidpunct}
{\mcitedefaultendpunct}{\mcitedefaultseppunct}\relax
\EndOfBibitem
\bibitem[G{\"u}nther and Jensen(2006)]{gunther2006multiphase}
A.~G{\"u}nther and K.~F. Jensen, \emph{Lab Chip}, 2006, \textbf{6},
  1487--1503\relax
\mciteBstWouldAddEndPuncttrue
\mciteSetBstMidEndSepPunct{\mcitedefaultmidpunct}
{\mcitedefaultendpunct}{\mcitedefaultseppunct}\relax
\EndOfBibitem
\bibitem[Assmann and von Rohr(2011)]{cep_assmann_2011}
N.~Assmann and P.~R. von Rohr, \emph{Chem. Eng. Process}, 2011, \textbf{50},
  822--827\relax
\mciteBstWouldAddEndPuncttrue
\mciteSetBstMidEndSepPunct{\mcitedefaultmidpunct}
{\mcitedefaultendpunct}{\mcitedefaultseppunct}\relax
\EndOfBibitem
\bibitem[{\L}adosz \emph{et~al.}(2016){\L}adosz, Rigger, and von
  Rohr]{mn_ladosz_2016}
A.~{\L}adosz, E.~Rigger and P.~R. von Rohr, \emph{Microfluid. Nanofluid}, 2016,
  \textbf{20}, 1--14\relax
\mciteBstWouldAddEndPuncttrue
\mciteSetBstMidEndSepPunct{\mcitedefaultmidpunct}
{\mcitedefaultendpunct}{\mcitedefaultseppunct}\relax
\EndOfBibitem
\bibitem[Heil \emph{et~al.}(2008)Heil, Hazel, and Smith]{heil2008mechanics}
M.~Heil, A.~L. Hazel and J.~A. Smith, \emph{Resp. Physiol. Neuro.}, 2008,
  \textbf{163}, 214--221\relax
\mciteBstWouldAddEndPuncttrue
\mciteSetBstMidEndSepPunct{\mcitedefaultmidpunct}
{\mcitedefaultendpunct}{\mcitedefaultseppunct}\relax
\EndOfBibitem
\bibitem[Kamm and Schroter(1989)]{kamm1989airway}
R.~Kamm and R.~Schroter, \emph{Resp. Phys.}, 1989, \textbf{75}, 141--156\relax
\mciteBstWouldAddEndPuncttrue
\mciteSetBstMidEndSepPunct{\mcitedefaultmidpunct}
{\mcitedefaultendpunct}{\mcitedefaultseppunct}\relax
\EndOfBibitem
\bibitem[Duclaux \emph{et~al.}(2006)Duclaux, Clanet, and
  Qu{\'e}r{\'e}]{duclaux2006effects}
V.~Duclaux, C.~Clanet and D.~Qu{\'e}r{\'e}, \emph{J. Fluid Mech.}, 2006,
  \textbf{556}, 217--226\relax
\mciteBstWouldAddEndPuncttrue
\mciteSetBstMidEndSepPunct{\mcitedefaultmidpunct}
{\mcitedefaultendpunct}{\mcitedefaultseppunct}\relax
\EndOfBibitem
\bibitem[Dietze and Ruyer-Quil(2015)]{dietze2015films}
G.~F. Dietze and C.~Ruyer-Quil, \emph{J. Fluid. Mech}, 2015, \textbf{762},
  68--109\relax
\mciteBstWouldAddEndPuncttrue
\mciteSetBstMidEndSepPunct{\mcitedefaultmidpunct}
{\mcitedefaultendpunct}{\mcitedefaultseppunct}\relax
\EndOfBibitem
\bibitem[Foroughi \emph{et~al.}(2012)Foroughi, Abbasi, Das, and
  Kawaji]{foroughi2012immiscible}
H.~Foroughi, A.~Abbasi, K.~S. Das and M.~Kawaji, \emph{Phys. Rev. E}, 2012,
  \textbf{85}, 026309\relax
\mciteBstWouldAddEndPuncttrue
\mciteSetBstMidEndSepPunct{\mcitedefaultmidpunct}
{\mcitedefaultendpunct}{\mcitedefaultseppunct}\relax
\EndOfBibitem
\bibitem[Gauglitz and Radke(1988)]{gauglitz1988extended}
P.~Gauglitz and C.~Radke, \emph{Chem. Eng. Sci.}, 1988, \textbf{43},
  1457--1465\relax
\mciteBstWouldAddEndPuncttrue
\mciteSetBstMidEndSepPunct{\mcitedefaultmidpunct}
{\mcitedefaultendpunct}{\mcitedefaultseppunct}\relax
\EndOfBibitem
\bibitem[Cassidy \emph{et~al.}(1999)Cassidy, Halpern, Ressler, and
  Grotberg]{cassidy1999surfactant}
K.~Cassidy, D.~Halpern, B.~Ressler and J.~Grotberg, \emph{J. Appl. Physiol},
  1999, \textbf{87}, 415--427\relax
\mciteBstWouldAddEndPuncttrue
\mciteSetBstMidEndSepPunct{\mcitedefaultmidpunct}
{\mcitedefaultendpunct}{\mcitedefaultseppunct}\relax
\EndOfBibitem
\bibitem[Burger and Macklem(1968)]{jap_burger_1968}
E.~Burger and P.~Macklem, \emph{J. Appl. Phys.}, 1968, \textbf{25},
  139--148\relax
\mciteBstWouldAddEndPuncttrue
\mciteSetBstMidEndSepPunct{\mcitedefaultmidpunct}
{\mcitedefaultendpunct}{\mcitedefaultseppunct}\relax
\EndOfBibitem
\bibitem[Hugues \emph{et~al.}(1970)Hugues, Rosenzweig, and
  Kivitz]{jap_hugues_1970}
J.~Hugues, D.~Rosenzweig and P.~Kivitz, \emph{J. Appl. Physiol.}, 1970,
  \textbf{29}, 340--344\relax
\mciteBstWouldAddEndPuncttrue
\mciteSetBstMidEndSepPunct{\mcitedefaultmidpunct}
{\mcitedefaultendpunct}{\mcitedefaultseppunct}\relax
\EndOfBibitem
\bibitem[Rogers(2007)]{rc_rogers_2007}
D.~Rogers, \emph{Resp. Care}, 2007, \textbf{52}, 1134--1149\relax
\mciteBstWouldAddEndPuncttrue
\mciteSetBstMidEndSepPunct{\mcitedefaultmidpunct}
{\mcitedefaultendpunct}{\mcitedefaultseppunct}\relax
\EndOfBibitem
\bibitem[Griese \emph{et~al.}(1997)Griese, Birrer, and
  Demirsoy]{griese1997pulmonary}
M.~Griese, P.~Birrer and A.~Demirsoy, \emph{Eur. Respir. J}, 1997, \textbf{10},
  1983--1988\relax
\mciteBstWouldAddEndPuncttrue
\mciteSetBstMidEndSepPunct{\mcitedefaultmidpunct}
{\mcitedefaultendpunct}{\mcitedefaultseppunct}\relax
\EndOfBibitem
\bibitem[Wright \emph{et~al.}(2000)Wright, Hockey, Enhorning, Strong, Reid,
  Holgate, Djukanovic, and Postle]{wright2000altered}
S.~M. Wright, P.~M. Hockey, G.~Enhorning, P.~Strong, K.~B. Reid, S.~T. Holgate,
  R.~Djukanovic and A.~D. Postle, \emph{J. Appl. Physiol}, 2000, \textbf{89},
  1283--1292\relax
\mciteBstWouldAddEndPuncttrue
\mciteSetBstMidEndSepPunct{\mcitedefaultmidpunct}
{\mcitedefaultendpunct}{\mcitedefaultseppunct}\relax
\EndOfBibitem
\bibitem[Hohlfeld \emph{et~al.}(2002)Hohlfeld\emph{et~al.}]{hohlfeld2002role}
J.~M. Hohlfeld \emph{et~al.}, \emph{Respir. Res}, 2002, \textbf{3}, 1--8\relax
\mciteBstWouldAddEndPuncttrue
\mciteSetBstMidEndSepPunct{\mcitedefaultmidpunct}
{\mcitedefaultendpunct}{\mcitedefaultseppunct}\relax
\EndOfBibitem
\bibitem[Grotberg(2011)]{pof_grotberg_2011}
J.~Grotberg, \emph{Phys. Fluids}, 2011, \textbf{23}, 021301\relax
\mciteBstWouldAddEndPuncttrue
\mciteSetBstMidEndSepPunct{\mcitedefaultmidpunct}
{\mcitedefaultendpunct}{\mcitedefaultseppunct}\relax
\EndOfBibitem
\bibitem[Engle \emph{et~al.}(2008)Engle\emph{et~al.}]{engle2008surfactant}
W.~A. Engle \emph{et~al.}, \emph{Pediatrics}, 2008, \textbf{121},
  419--432\relax
\mciteBstWouldAddEndPuncttrue
\mciteSetBstMidEndSepPunct{\mcitedefaultmidpunct}
{\mcitedefaultendpunct}{\mcitedefaultseppunct}\relax
\EndOfBibitem
\bibitem[Barber and Blaisdell(2010)]{barber2010respiratory}
M.~Barber and C.~J. Blaisdell, \emph{Am. J. Perinat}, 2010, \textbf{27},
  549--558\relax
\mciteBstWouldAddEndPuncttrue
\mciteSetBstMidEndSepPunct{\mcitedefaultmidpunct}
{\mcitedefaultendpunct}{\mcitedefaultseppunct}\relax
\EndOfBibitem
\bibitem[Espinosa and Kamm(1998)]{espinosa1998meniscus}
F.~Espinosa and R.~Kamm, \emph{J. Appl. Physiol}, 1998, \textbf{85},
  266--272\relax
\mciteBstWouldAddEndPuncttrue
\mciteSetBstMidEndSepPunct{\mcitedefaultmidpunct}
{\mcitedefaultendpunct}{\mcitedefaultseppunct}\relax
\EndOfBibitem
\bibitem[Stevens and Sinkin(2007)]{stevens2007surfactant}
T.~P. Stevens and R.~A. Sinkin, \emph{Chest}, 2007, \textbf{131},
  1577--1582\relax
\mciteBstWouldAddEndPuncttrue
\mciteSetBstMidEndSepPunct{\mcitedefaultmidpunct}
{\mcitedefaultendpunct}{\mcitedefaultseppunct}\relax
\EndOfBibitem
\bibitem[Halpern \emph{et~al.}(2008)Halpern, Fujioka, Takayama, and
  Grotberg]{halpern2008liquid}
D.~Halpern, H.~Fujioka, S.~Takayama and J.~B. Grotberg, \emph{Resp. Physiol.
  Neuro.}, 2008, \textbf{163}, 222--231\relax
\mciteBstWouldAddEndPuncttrue
\mciteSetBstMidEndSepPunct{\mcitedefaultmidpunct}
{\mcitedefaultendpunct}{\mcitedefaultseppunct}\relax
\EndOfBibitem
\bibitem[Van't~Veen \emph{et~al.}(1998)Van't~Veen, Wollmer, Nilsson, Gommers,
  Mouton, Kooij, and Lachmann]{acp_vantveen_1998}
A.~Van't~Veen, P.~Wollmer, L.~Nilsson, D.~Gommers, J.~Mouton, P.~Kooij and
  B.~Lachmann, \emph{ACP-Appl. Cardiopul. P}, 1998, \textbf{7}, 87--94\relax
\mciteBstWouldAddEndPuncttrue
\mciteSetBstMidEndSepPunct{\mcitedefaultmidpunct}
{\mcitedefaultendpunct}{\mcitedefaultseppunct}\relax
\EndOfBibitem
\bibitem[Nimmo \emph{et~al.}(2002)Nimmo, Carstairs, Patole, Whitehall,
  Davidson, and Vink]{cepp_nimmo_2002}
A.~Nimmo, J.~Carstairs, S.~Patole, J.~Whitehall, K.~Davidson and R.~Vink,
  \emph{Clin. Exp. Pharmocol. Physiol.}, 2002, \textbf{29}, 661\relax
\mciteBstWouldAddEndPuncttrue
\mciteSetBstMidEndSepPunct{\mcitedefaultmidpunct}
{\mcitedefaultendpunct}{\mcitedefaultseppunct}\relax
\EndOfBibitem
\bibitem[Aussillous and Qu{\'e}r{\'e}(2000)]{aussillous2000quick}
P.~Aussillous and D.~Qu{\'e}r{\'e}, \emph{Phys. Fluids}, 2000, \textbf{12},
  2367--2371\relax
\mciteBstWouldAddEndPuncttrue
\mciteSetBstMidEndSepPunct{\mcitedefaultmidpunct}
{\mcitedefaultendpunct}{\mcitedefaultseppunct}\relax
\EndOfBibitem
\bibitem[Bico and Qu{\'e}r{\'e}(2001)]{bico2001falling}
J.~Bico and D.~Qu{\'e}r{\'e}, \emph{J. Colloid. Interf. Sci}, 2001,
  \textbf{243}, 262--264\relax
\mciteBstWouldAddEndPuncttrue
\mciteSetBstMidEndSepPunct{\mcitedefaultmidpunct}
{\mcitedefaultendpunct}{\mcitedefaultseppunct}\relax
\EndOfBibitem
\bibitem[Fries \emph{et~al.}(2008)Fries, Trachsel, and Rudolf~von
  Rohr]{ijmf_fries_2008}
D.~Fries, F.~Trachsel and P.~Rudolf~von Rohr, \emph{Int. J. Mult. Flow}, 2008,
  \textbf{34}, 1108--1118\relax
\mciteBstWouldAddEndPuncttrue
\mciteSetBstMidEndSepPunct{\mcitedefaultmidpunct}
{\mcitedefaultendpunct}{\mcitedefaultseppunct}\relax
\EndOfBibitem
\bibitem[Zheng \emph{et~al.}(2009)Zheng, Fujioka, Bian, Torisawa, Huh,
  Takayama, and Grotberg]{zheng2009liquid}
Y.~Zheng, H.~Fujioka, S.~Bian, Y.~Torisawa, D.~Huh, S.~Takayama and
  J.~Grotberg, \emph{Phys. Fluids}, 2009, \textbf{21}, 071903\relax
\mciteBstWouldAddEndPuncttrue
\mciteSetBstMidEndSepPunct{\mcitedefaultmidpunct}
{\mcitedefaultendpunct}{\mcitedefaultseppunct}\relax
\EndOfBibitem
\bibitem[Song \emph{et~al.}(2011)Song, Baudoin, Manneville, and
  Baroud]{song2011air}
Y.~Song, M.~Baudoin, P.~Manneville and C.~N. Baroud, \emph{Med. Eng. Phys.},
  2011, \textbf{33}, 849--856\relax
\mciteBstWouldAddEndPuncttrue
\mciteSetBstMidEndSepPunct{\mcitedefaultmidpunct}
{\mcitedefaultendpunct}{\mcitedefaultseppunct}\relax
\EndOfBibitem
\bibitem[Baudoin \emph{et~al.}(2013)Baudoin, Song, Manneville, and
  Baroud]{baudoin2013airway}
M.~Baudoin, Y.~Song, P.~Manneville and C.~N. Baroud, \emph{Proc. Nat. Ac. Sci.
  USA}, 2013, \textbf{110}, 859--864\relax
\mciteBstWouldAddEndPuncttrue
\mciteSetBstMidEndSepPunct{\mcitedefaultmidpunct}
{\mcitedefaultendpunct}{\mcitedefaultseppunct}\relax
\EndOfBibitem
\bibitem[Hu \emph{et~al.}(2015)Hu, Bian, Grotberg, Filoche, White, Takayama,
  and Grotberg]{hu2015microfluidic}
Y.~Hu, S.~Bian, J.~Grotberg, M.~Filoche, J.~White, S.~Takayama and J.~B.
  Grotberg, \emph{Biomicrofluidics}, 2015, \textbf{9}, 044119\relax
\mciteBstWouldAddEndPuncttrue
\mciteSetBstMidEndSepPunct{\mcitedefaultmidpunct}
{\mcitedefaultendpunct}{\mcitedefaultseppunct}\relax
\EndOfBibitem
\bibitem[Fujioka and Grotberg(2005)]{pof_fujioka_2005}
H.~Fujioka and J.~Grotberg, \emph{Phys Fluids}, 2005, \textbf{17}, 082102\relax
\mciteBstWouldAddEndPuncttrue
\mciteSetBstMidEndSepPunct{\mcitedefaultmidpunct}
{\mcitedefaultendpunct}{\mcitedefaultseppunct}\relax
\EndOfBibitem
\bibitem[Campana \emph{et~al.}(2007)Campana, Ubal, Giavedoni, and
  Saita]{campana2007stability}
D.~M. Campana, S.~Ubal, M.~D. Giavedoni and F.~A. Saita, \emph{Ind. Eng. Chem.
  Res.}, 2007, \textbf{46}, 1803--1809\relax
\mciteBstWouldAddEndPuncttrue
\mciteSetBstMidEndSepPunct{\mcitedefaultmidpunct}
{\mcitedefaultendpunct}{\mcitedefaultseppunct}\relax
\EndOfBibitem
\bibitem[Fujioka \emph{et~al.}(2008)Fujioka, Takayama, and
  Grotberg]{fujioka2008unsteady}
H.~Fujioka, S.~Takayama and J.~B. Grotberg, \emph{Phys Fluids}, 2008,
  \textbf{20}, 062104\relax
\mciteBstWouldAddEndPuncttrue
\mciteSetBstMidEndSepPunct{\mcitedefaultmidpunct}
{\mcitedefaultendpunct}{\mcitedefaultseppunct}\relax
\EndOfBibitem
\bibitem[Ubal \emph{et~al.}(2008)Ubal, Campana, Giavedoni, and
  Saita]{ubal2008stability}
S.~Ubal, D.~M. Campana, M.~D. Giavedoni and F.~A. Saita, \emph{Ind. Eng. Chem.
  Res}, 2008, \textbf{47}, 6307--6315\relax
\mciteBstWouldAddEndPuncttrue
\mciteSetBstMidEndSepPunct{\mcitedefaultmidpunct}
{\mcitedefaultendpunct}{\mcitedefaultseppunct}\relax
\EndOfBibitem
\bibitem[Hassan \emph{et~al.}(2011)Hassan, Uzgoren, Fujioka, Grotberg, and
  Shyy]{hassan2011adaptive}
E.~A. Hassan, E.~Uzgoren, H.~Fujioka, J.~B. Grotberg and W.~Shyy, \emph{Int. J.
  Num. Meth. Fluids}, 2011, \textbf{67}, 1373--1392\relax
\mciteBstWouldAddEndPuncttrue
\mciteSetBstMidEndSepPunct{\mcitedefaultmidpunct}
{\mcitedefaultendpunct}{\mcitedefaultseppunct}\relax
\EndOfBibitem
\bibitem[Ratulowski and Chang(1989)]{pof_ratulowski_1989}
J.~Ratulowski and H.-C. Chang, \emph{Phys. Fluids A.}, 1989,  1642\relax
\mciteBstWouldAddEndPuncttrue
\mciteSetBstMidEndSepPunct{\mcitedefaultmidpunct}
{\mcitedefaultendpunct}{\mcitedefaultseppunct}\relax
\EndOfBibitem
\bibitem[Jensen(2000)]{jcis_jensen_2000}
O.~Jensen, \emph{J. Coll. Interf. Sci.}, 2000, \textbf{221}, 38--49\relax
\mciteBstWouldAddEndPuncttrue
\mciteSetBstMidEndSepPunct{\mcitedefaultmidpunct}
{\mcitedefaultendpunct}{\mcitedefaultseppunct}\relax
\EndOfBibitem
\bibitem[Kreutzer \emph{et~al.}(2005)Kreutzer, Kapteijn, Moulijn, Fleijn, and
  Heiszwolf]{a_kreutzer_2005}
M.~Kreutzer, F.~Kapteijn, J.~Moulijn, C.~Fleijn and J.~Heiszwolf, \emph{AIChE
  J.}, 2005, \textbf{51}, 2428--2440\relax
\mciteBstWouldAddEndPuncttrue
\mciteSetBstMidEndSepPunct{\mcitedefaultmidpunct}
{\mcitedefaultendpunct}{\mcitedefaultseppunct}\relax
\EndOfBibitem
\bibitem[Warnier \emph{et~al.}(2010)Warnier, De~Croon, Rebrov, and
  Shouten]{mn_warnier_2010}
M.~Warnier, M.~De~Croon, E.~Rebrov and J.~Shouten, \emph{Microfluid.
  Nanofluid.}, 2010, \textbf{8}, 33--45\relax
\mciteBstWouldAddEndPuncttrue
\mciteSetBstMidEndSepPunct{\mcitedefaultmidpunct}
{\mcitedefaultendpunct}{\mcitedefaultseppunct}\relax
\EndOfBibitem
\bibitem[Filoche \emph{et~al.}(2015)Filoche, Tai, and
  Grotberg]{filoche2015three}
M.~Filoche, C.-F. Tai and J.~B. Grotberg, \emph{Proc. Nat. Ac. Sci. USA}, 2015,
  \textbf{112}, 9287--9292\relax
\mciteBstWouldAddEndPuncttrue
\mciteSetBstMidEndSepPunct{\mcitedefaultmidpunct}
{\mcitedefaultendpunct}{\mcitedefaultseppunct}\relax
\EndOfBibitem
\bibitem[Stewart and Jensen(2015)]{stewart2015patterns}
P.~S. Stewart and O.~E. Jensen, \emph{J. R. Soc. Interface}, 2015, \textbf{12},
  20150523\relax
\mciteBstWouldAddEndPuncttrue
\mciteSetBstMidEndSepPunct{\mcitedefaultmidpunct}
{\mcitedefaultendpunct}{\mcitedefaultseppunct}\relax
\EndOfBibitem
\bibitem[Weibel(1984)]{weibel1984pathway}
E.~R. Weibel, \emph{The pathway for oxygen: structure and function in the
  mammalian respiratory system}, Harvard University Press, 1984\relax
\mciteBstWouldAddEndPuncttrue
\mciteSetBstMidEndSepPunct{\mcitedefaultmidpunct}
{\mcitedefaultendpunct}{\mcitedefaultseppunct}\relax
\EndOfBibitem
\bibitem[West \emph{et~al.}(1997)West, Brown, and Enquist]{west1997general}
G.~B. West, J.~H. Brown and B.~J. Enquist, \emph{Science}, 1997, \textbf{276},
  122--126\relax
\mciteBstWouldAddEndPuncttrue
\mciteSetBstMidEndSepPunct{\mcitedefaultmidpunct}
{\mcitedefaultendpunct}{\mcitedefaultseppunct}\relax
\EndOfBibitem
\bibitem[Mauroy \emph{et~al.}(2004)Mauroy, Filoche, Weibel, and
  Sapoval]{mauroy2004optimal}
B.~Mauroy, M.~Filoche, E.~Weibel and B.~Sapoval, \emph{Nature}, 2004,
  \textbf{427}, 633--636\relax
\mciteBstWouldAddEndPuncttrue
\mciteSetBstMidEndSepPunct{\mcitedefaultmidpunct}
{\mcitedefaultendpunct}{\mcitedefaultseppunct}\relax
\EndOfBibitem
\bibitem[Pedley(1977)]{pedley1977pulmonary}
T.~Pedley, \emph{Annu. Rev. Fluid. Mech}, 1977, \textbf{9}, 229--274\relax
\mciteBstWouldAddEndPuncttrue
\mciteSetBstMidEndSepPunct{\mcitedefaultmidpunct}
{\mcitedefaultendpunct}{\mcitedefaultseppunct}\relax
\EndOfBibitem
\bibitem[West \emph{et~al.}(1986)West, Bhargava, and
  Goldberger]{west1986beyond}
B.~J. West, V.~Bhargava and A.~Goldberger, \emph{J. Appl. Physiol}, 1986,
  \textbf{60}, 1089--1097\relax
\mciteBstWouldAddEndPuncttrue
\mciteSetBstMidEndSepPunct{\mcitedefaultmidpunct}
{\mcitedefaultendpunct}{\mcitedefaultseppunct}\relax
\EndOfBibitem
\bibitem[Association
  \emph{et~al.}(1963)Association\emph{et~al.}]{glycerine1963physical}
G.~P. Association \emph{et~al.}, \emph{Physical properties of glycerine and its
  solutions}, Glycerine Producers' Association, 1963\relax
\mciteBstWouldAddEndPuncttrue
\mciteSetBstMidEndSepPunct{\mcitedefaultmidpunct}
{\mcitedefaultendpunct}{\mcitedefaultseppunct}\relax
\EndOfBibitem
\bibitem[Bretherton(1961)]{bretherton1961motion}
F.~Bretherton, \emph{J. Fluid. Mech}, 1961, \textbf{10}, 166--188\relax
\mciteBstWouldAddEndPuncttrue
\mciteSetBstMidEndSepPunct{\mcitedefaultmidpunct}
{\mcitedefaultendpunct}{\mcitedefaultseppunct}\relax
\EndOfBibitem
\bibitem[Chebbi(2003)]{chebbi2003deformation}
R.~Chebbi, \emph{J. Colloid. Interf. Sci}, 2003, \textbf{265}, 166--173\relax
\mciteBstWouldAddEndPuncttrue
\mciteSetBstMidEndSepPunct{\mcitedefaultmidpunct}
{\mcitedefaultendpunct}{\mcitedefaultseppunct}\relax
\EndOfBibitem
\bibitem[Widdicombe(2002)]{widdicombe2002regulation}
J.~Widdicombe, \emph{J. Anatomy}, 2002, \textbf{201}, 313--318\relax
\mciteBstWouldAddEndPuncttrue
\mciteSetBstMidEndSepPunct{\mcitedefaultmidpunct}
{\mcitedefaultendpunct}{\mcitedefaultseppunct}\relax
\EndOfBibitem
\bibitem[King and Rubin(1996)]{King}
M.~King and B.~Rubin, \emph{Acute respiratory failure in chronic obstructive
  pulmonarydisease}, 1996,  chapter 13 : Mucus physiology and pathophysiology :
  391–411\relax
\mciteBstWouldAddEndPuncttrue
\mciteSetBstMidEndSepPunct{\mcitedefaultmidpunct}
{\mcitedefaultendpunct}{\mcitedefaultseppunct}\relax
\EndOfBibitem
\bibitem[Grotberg(1994)]{grotbergpulmonary}
J.~Grotberg, \emph{Ann. Rev. Fluid Mech.}, 1994, \textbf{26}, 529--571\relax
\mciteBstWouldAddEndPuncttrue
\mciteSetBstMidEndSepPunct{\mcitedefaultmidpunct}
{\mcitedefaultendpunct}{\mcitedefaultseppunct}\relax
\EndOfBibitem
\bibitem[Heil and White(2002)]{heil2002airway}
M.~Heil and J.~P. White, \emph{J. of Fluid Mech.}, 2002, \textbf{462},
  79--109\relax
\mciteBstWouldAddEndPuncttrue
\mciteSetBstMidEndSepPunct{\mcitedefaultmidpunct}
{\mcitedefaultendpunct}{\mcitedefaultseppunct}\relax
\EndOfBibitem
\bibitem[Marieb and Hoehn(2014)]{marieb2014anatomie}
E.~Marieb and K.~Hoehn, \emph{Anatomie et physiologie humaines}, Pearson
  Education France, 2014\relax
\mciteBstWouldAddEndPuncttrue
\mciteSetBstMidEndSepPunct{\mcitedefaultmidpunct}
{\mcitedefaultendpunct}{\mcitedefaultseppunct}\relax
\EndOfBibitem
\bibitem[Zamankhan \emph{et~al.}(2012)Zamankhan, Helenbrook, Takayama, and
  Grotberg]{zamankhan2012steady}
P.~Zamankhan, B.~T. Helenbrook, S.~Takayama and J.~B. Grotberg, \emph{J. Fluid.
  Mech}, 2012, \textbf{705}, 258--279\relax
\mciteBstWouldAddEndPuncttrue
\mciteSetBstMidEndSepPunct{\mcitedefaultmidpunct}
{\mcitedefaultendpunct}{\mcitedefaultseppunct}\relax
\EndOfBibitem
\end{mcitethebibliography}
\bibliographystyle{rsc} 
}

\end{document}